# Research on Information Security Enhancement Approaches and the Applications on HCI Systems

Yang Yang, Zhiyan Li
Tianjin Electronic Information College
City of Tianjin, 300350, China

*Abstract*—With rapid development of computer techniques, the human computer interaction scenarios are becoming more and more frequent. The development history of the human-computer interaction is from a person to adapt to the computer to the computer and continually adapt to the rapid development. Facing the process of human-computer interaction, information system daily operation to produce huge amounts of data, how to ensure human-computer interaction interface clear, generated data safe and reliable, has become a problem to be solved in the world of information. To deal with the challenging, we propose the information security enhancement approaches and the core applications on HCI systems. Through reviewing the other state-of-the-art methods, we propose the data encryption system to deal with the issues that uses mixed encryption system to make full use of the symmetric cipher algorithm encryption speed and encryption intensity is high while the encryption of large amounts of data efficiently. Our method could enhance the general safety of the HCI system, the experimental result verities the feasibility and general robustness of our approach.

## I. INTRODUCTION

Human-computer interaction (HCI) is a human, computer technology and the mutual influence between them and the user interface is the medium of transmission, the exchange of information between people and computer and dialogue interface that is an important part of a computer system. The human-computer interaction and user interface are closely linked, but it is two different concepts: the former emphasizes the technology and model, which is a critical part of computer. The development history of the human-computer interaction is from a person to adapt to the computer to the computer and continually adapt to the rapid development. It has experienced several stages as the follows [1-2]. (1) In the early stages of manual work, the characteristics of the interaction was by the designer to use the computer, they adopt the method of manual operation and rely on machines to now seem very clumsy. (2) Job control language and interactive command language stage. The characteristics of this stage is the main user of computer, the programmer can use batch jobs or the interactive command language way of dealing with a computer, although want to remember many commands and expertly knocked at keyboard, but has a convenient means to debug programs available, and understand computer implementation. (3) Phase of graphical user interface. The main characteristic of the GUI is desktop metaphor, direct manipulation and "wysiwyg". Because the GUI clear and easy to learn, to reduce the typing, implement the "standardized", in fact, so that don't understand, ordinary users also can skillfully use the computer to exploit the user population, its emergence has the information industry got unprecedented development. (4) The emergence of the web user interface. To hypertext markup language HTML and the hypertext transfer protocol HTTP as the main basis of the web browser is a representative of the network user interface. The WWW network formed by it has become Internet backbone. (5) Multi-channel, the multimedia intelligent human-computer interaction stage. Represented by virtual reality of computer system of the anthropomorphic and represented by handheld computers, smart phones, computer miniaturization, portable embedded, are the two important development trend of the current computer [3].

Besides the HCI system itself, state-of-the-art other related techniques are considered while constructing the novel idea. Image processing and computer vision techniques could assist obtaining the figures and extract the necessary features [4-7], machine learning can help to mining the data and obtain the information for control [8-10], data security and cloud storage techniques could enhance the robustness and efficiency of the systems [11-14]. Among mentioned issues, the data security concern is recently well studied for the appearance of the hacking phenomenon.

From the man-machine interactive data classification, it can be roughly divided into: structured data, unstructured data and semi-structured data. Structured data refers to the way by two-dimensional table structure logic expression data, such as data stored in the database; Unstructured data is not convenient to use a two-dimensional logical table performance data, such as all formats of office documents, such as text, images, audio, all kinds of statements data; Semi-structured data is structured data in the performance form, but the structure changed a lot, such as store employee resumes, usually the semi-structured data into the structured data for processing. In the daily office environment in the process of the human-computer interaction, structured and the semi-structured data through database for the centralized storage, that mostly adopt protective units unified security policy, its security is relatively high. Unstructured data are scattered on the personal computer in enterprise network, and faces the risk of each different, each person's calculation of the human-computer interaction interface protective measures is uneven, its security cannot be guaranteed, so unstructured data security is the focus of the current data security.

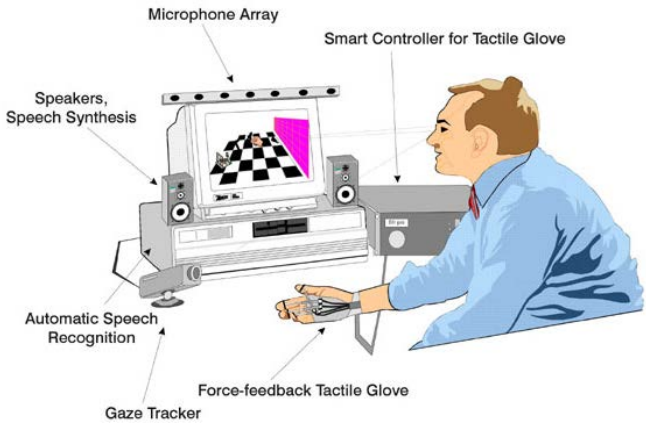

Fig. 1. The Modern HCI System Demonstration

As shown in the figure one, we illustrate the modern HCI system architecture. In this paper, we will conduct theoretical analysis on the information security enhancement approaches and the applications on HCI systems. The reminding of the paper is organized as the follows. In the section two, we review the state-of-the-art information security enhancement methods and propose the revise paradigm for the data security protection model. In the section three, we apply the proposed model on the HCI systems and find out the optimal combination. Later, in the section four, we simulate the performance of the data enhancement algorithm with empirical analysis. Finally, we conclude the work and schedule our future research.

## II. INFORMATION SECURITY ENHANCEMENT MODELS

### A. Information Encryption Technology Review

In the process of informatization, information security problem is very important and essential problem. However, in the information security technology, the password techniques play core role. The characteristics of the information security requirements to meet the following parts, namely: the integrity, confidentiality, and the controllability. Through the password techniques can be used to protect information from the illegal access by authorized users. At present, according to the use of key species key system can be divided into the following two kinds: first, the private key password; second, the public key password. The private key password has the very distinctive characteristics, that namely: we can now be done using a key encryption, decryption both work. To some extent, the safety factor of the private key is very high and then the key system compromised [15-16]. The prominent characteristics of public key cryptography are the use of different keys to encrypt and decrypt the two works. PSA public key cryptosystem, knapsack public key cryptosystem, the influence is very big. But there exists a deficiency, public key cryptography that is public key cryptography algorithm is complex and decryption speed is very slow. So, it will make private key password and public key cryptography to combine the two phase mixed classified system into a kind of encryption method which is relatively commonly used. By using the end-to-end encryption can be according to the application to manage keys, and can according to the requirement of the communication object to change the encryption keys in a reasonable manner. It is well known that link encryption to protect whole link communication system, however with end-to-end way can protect the whole network system, thus, in the field of development in the future, end-to-end encryption application prospect is very broad.

Select the routing in the process of the computer network occupies the very important position. Refers to the so-called routing algorithm is to be able to realize routing function of an algorithm. Typically, the routing algorithms can be seen as a communication protocol in network layer. By studying the routing algorithm, it not only can make to improve the core transmission efficiency of communication subnet and also can ensure the reliability of communication subnet. The figure two shows the corresponding scenarios.

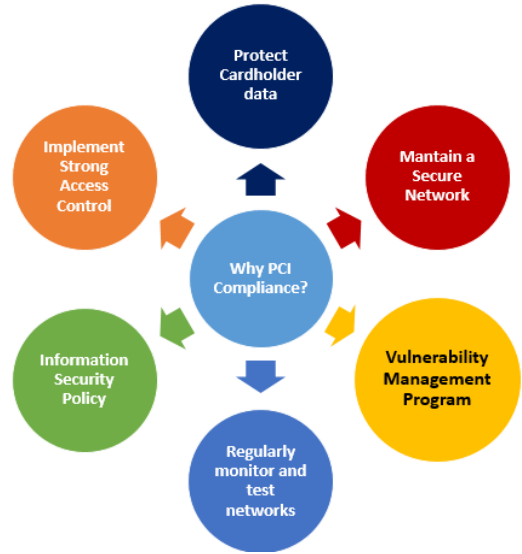

Fig. 2. The Information Encryption Technology Application Scenarios

Monitoring type firewall technology is the product of the rapid development of science and technology. Monitoring the function of the type of firewall technology has gone far beyond the traditional function of firewall technology. Monitoring type of firewall technology can active, real-time monitoring data of each layer, and the analysis of these data as the foundation, accurate judgment whether there are illegal invasion in each layer at the same time in various application servers and other network nodes are installed with distributed detectors. By installing distributed detector can detect whether there are external network attack that can also detect whether there are external network damage internal system to from perspective of security generations in front of the incomparable advantage.

### B. The Proposed Security Enhancement Algorithm

Data encryption technology is based on the encryption algorithm as the core. According to the cipher algorithm used by the encryption key and decryption key are same, whether by encryption, decryption process is deduced from the factors such as the process, the password system can be divided into symmetric key encryption system and primary asymmetric key encryption system [17-20].

Data encryption system uses the mixed encryption system, make full use of the symmetric cipher algorithm encryption speed and encryption intensity is high, the encryption of large amounts of data efficiently. Use of public key cryptography algorithm encryption intensity is high, the key is convenient for management, realize the clear key is encrypted, the key to make up for traditional password algorithm is not convenient to transfer the weakness as a combination of these, in order to achieve the security of data transmission. Accordingly, the data security lifecycle for our methodology is as below.

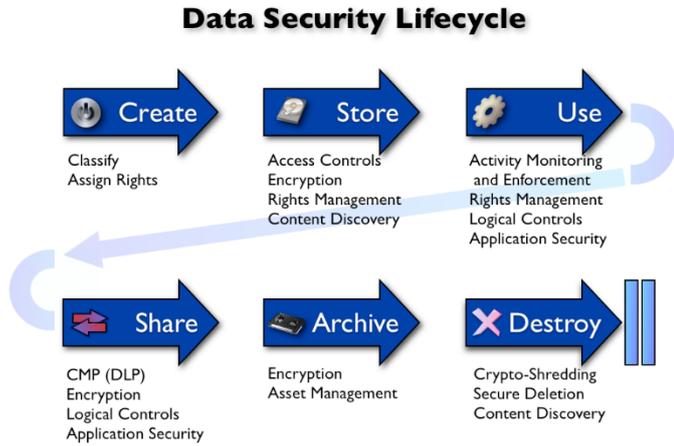

Fig. 3. The Data Security Lifecycle for Our Methodology

Symmetric key encryption system is also called the core encryption system. In this kind of encryption method, try to use the same key to encrypt and decrypt information, or from a key can be easily deduced another key. For the later discussion, we firstly define the proclaimed in writing, the random invertible matrix and encrypted document in formula 1~3.

$$M = (m_1, m_2, \cdots, m_n) \quad (1)$$

$$K = (k_{ij})_{n \times n} \quad (2)$$

$$C = KM = (c_1, c_2, \cdots, c_n) \quad (3)$$

Advantages of this mechanism is simple, communication both sides need to exchange each other keys, the encryption process can be realized. The algorithm of the symmetric encryption system to achieve speed, software implementation speed reached trillions or the dozens of megabit per second. However, symmetric cipher algorithm exist the following problems: key can't secretly distribution; Lack of automatically detecting the key leaked ability; Assumption in the network each use different keys to the customer, so the total number of key increased with the increase of users and fast that can't solve the problem of confirmation. The improvement program of the encryption algorithm made full use of the diversity of the matrix operations, but there is a certain difficulty to choose the appropriate encryption matrix, and the encryption matrix of a single, easy to break to make up for this, using the algorithm of tensor product further improvement: it is using small invertible matrix to construct a larger order of the encryption matrix. To illustrate encryption matrix tensor product structure, we give the following basic results.

Suppose the $m$ represents the positive integer, we have the corresponding non-negative integer set as follows.

$$E(m) = \{a \in Z \mid 0 \le a \le 2^m\} \quad (4)$$

For any second-order reversible phalanx, its tensor product can be expressed as the follows.

$$A = \otimes \sum_{i=1}^{n} A_i \quad (5)$$

Based on this, we could get the digital signature. Digital signature with symmetric algorithm implementation, if the third party certification, more trouble and if the public key encryption algorithm, can avoid the trouble. The public key encryption algorithm is adopted to realize digital signature and authentication process is as the follows. (1) Sender first to a message using open one-way function transform, get a digital signature, and then the digital signature to make use of the private key encrypted with the attached to the message after send out. (2) The receiver will be with the sender's public key to decrypt the digital signature for the exchange, proclaimed in writing by a digital signature. The sender's public key can be a trusted technology certification center the administrative body of the recipient to get the clear through one-way function to calculate, also get a digital signature, and then compare the two digital signature, if the same, is proved that signature is valid, or otherwise is invalid.

As with the Java, we discuss the implementation steps as the follows. To realize digital signature, including emergence of a key generation, signature, signature verification and the process of the management of the key, we can use the Java classes defined in the create RSA public key and private key. (1) Key generation: Java KeyPairGnerator class provides the methods to create a key pair, so that is used in the asymmetric encryption, key to create good encapsulation in KeyPair type of object, after KeyPair class provides access to the public and private keys. (2) Using the SHA to generate digital abstract: from the private key file prikey.txt, in order to the information in the file signature, signature will be stored in file. Javax.mail. Security package the MessageDigest class provides a way to generate digital paper. (3) Using the RSA digital signature: after the operation, the generated Numbers based on the RSA signature certification. (4) Verify the digital signature with the public key: when the receiver to receive after departure from anything ciphertext and the signature, to test and verify, the premise is the receiver with the sender's public key. Use the received ciphertext and its signature file verification, to ensure that has not been altered the original text, and is, indeed, from the sender. Additionally, we could also use the initverfy() method to input the public key.

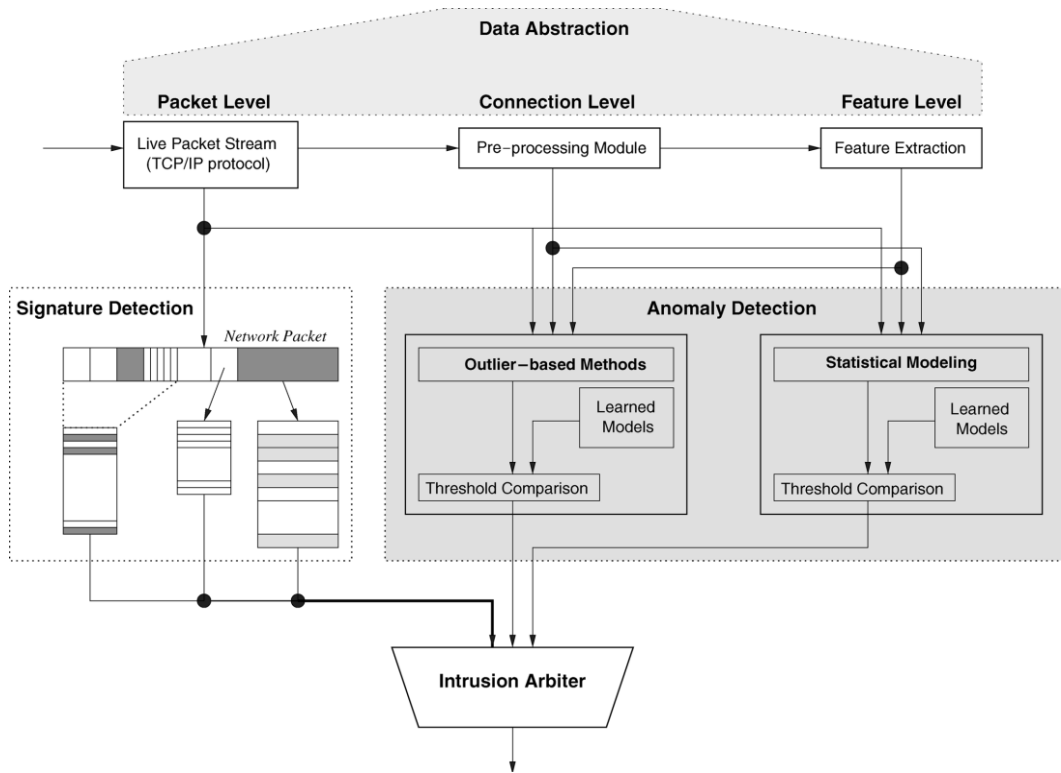

Fig. 4. The Systematic Description of the Proposed Security Enhancement Architecture

### III. HCI SYSTEMS AND SECURITY

Interactive data refers to the electronic document, ontology is refers to the electronic document itself. So-called human-computer interaction data security of ontology is to point to: document production, processing, storage, transmission, exchange and destroy the whole life cycle of its own security, primarily to ensure that the documents confidentiality, integrity and non-repudiation, to ensure the safety of the file itself in the process of human-computer interaction.

Integrity refers to the consistency and accuracy of the data, to prevent the data in process of human-computer interaction is used by the grantee or tampered with, resist the threat of illegal operators for data, ensure authenticity of data. Confidentiality, it is to point to in the process of human-computer interaction, electronic files are not use or disclose the grantee, entity or process characteristics. Non-repudiation, maintain data non-repudiation evidence, on the one hand, to prevent the author denies the source of the data, on the other hand to protect the copyright data from the holder. Based on the prior discussion on the information security paradigm, we apply it on modern human-computer interaction systems.

Process identification is mainly analyzed the process, to determine whether a process is a controlled process. As long as it is a controlled process its operating documents are encrypted. As setting the winword controlled process, the doc files, are mandatory and transparently encryption. Only controlled by the same token, the process to have the right to access to clear, the other uncontrolled process during a visit to doc file can only get the ciphertext. At the same time, the identification of a controlled process, by unique identification of the fingerprint identification technology process, prevent malware disguised as a controlled process to steal data. It is to have transparency and mandatory transparent encryption key characteristics. Transparency refers to the user in the process of operation do not change the file using habit with the encryption operation process in the background automatically.

Format protection is based on the format conversion of the original data is according to the certain standard of editing processing as can make the different applications of data are characterized by a specific rule of data object, and through independent reader or embedded in the application software of control the data access control. Format based on the format conversion technology mainly by means of digital copyright protection technology, the protected data access management. Ontology of primary human-computer interaction data safety protection, at present, the development of the overall trend is to ensure data confidentiality, integrity, and non-repudiation for the principle, aiming at safety as far as possible reduce the human-computer interaction process, transparent encryption, format of ontology safety protection to protect the data in the future that still occupy the important position.

Based on this, we could propose the further developmental trend for HCI system security enhancement could be generally summarized as the follows. (1) Active security encryption is ontology of auxiliary measures, auxiliary measures as human-computer interaction data safety of ontology as secrecy is not high, large work volume on human-computer interaction, and classified information scenarios is the main development focus.

(2) Transparent encryption is the core of the current and future development of data security and encryption as the leading factor, to drive the layer to reduce the security risks of human-computer interaction. (3) File path tracking, it is the file from creation, use and destruction of the entire life cycle track and audit, to data ontology based closed-loop controlled safety.

## IV. EXPERIMENT AND VERIFICATION

To test the robustness of our proposed method, in this section, we conduct experiment on the methods. In order to realize the network communication system in the encrypted file transmission between networks, this system uses the Windows socket programming on the Windows XP operating system platform, based on the TCP/IP protocol with the use of Java socket programming to achieve the file transfer. At the same time, we conduct comparison experiment with other ones. The experiment result is shown in figure 5.

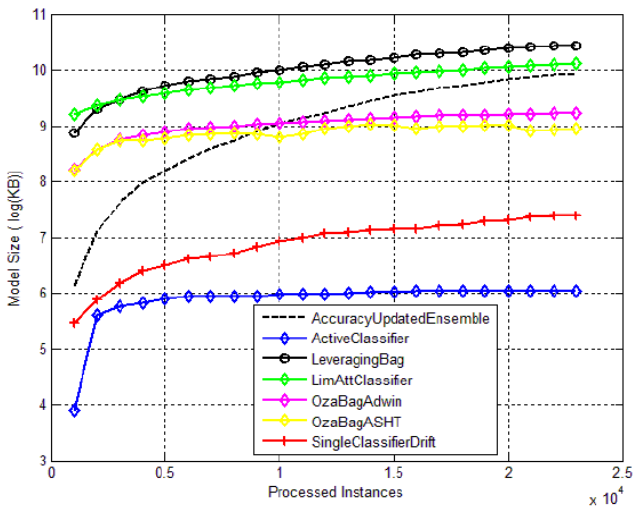

Fig. 5. The Simulation Result on the Security Enhancement Performance

## V. CONCLUSION

This paper proposes the novel perspectives on information security enhancement approaches and the applications on HCI systems. Data as a product of human-computer interaction, it is operating the source of power information system, in the daily operation of enterprises, state machine plays an irreplaceable role, in production, processing, storage, and exchange data in the process of life cycle, security threats and risks resulting from the human-computer interaction is everywhere, so the leakage of state secret, commercial secrets happens quite often. To enhance the data and information security level for the HCI systems, we propose the data encryption system to deal with the issues that uses the mixed encryption system, make full use of symmetric cipher algorithm encryption speed and encryption intensity is high while the encryption of large amounts of data efficiently. Through the experimental analysis, the robustness and effectiveness of the method is verified.

REFERENCES

[1] Ebert, Achim, Nahum D. Gershon, and Gerrit C. van der Veer. "Human-computer interaction." KI-Künstliche Intelligenz 26.2 (2012): 121-126.

[2] Rautaray, Siddharth S., and Anupam Agrawal. "Vision based hand gesture recognition for human computer interaction: a survey." Artificial Intelligence Review 43.1 (2015): 1-54.

[3] Barea, Rafael, et al. "EOG-based eye movements codification for human computer interaction." Expert Systems with Applications 39.3 (2012): 2677-2683.

[4] Mazzoli, Alida, and Orlando Favoni. "Particle size, size distribution and morphological evaluation of airborne dust particles of diverse woods by Scanning Electron Microscopy and image processing program." Powder Technology 225 (2012): 65-71.

[5] Zelelew, H. M., et al. "An improved image processing technique for asphalt concrete X-ray CT images." Road Materials and Pavement Design 14.2 (2013): 341-359.

[6] Wang, Haoxiang, and Jingbin Wang. "An effective image representation method using kernel classification." Tools with Artificial Intelligence (ICTAI), 2014 IEEE 26th International Conference on. IEEE, 2014.

[7] Kim, Sung-Wan, and Nam-Sik Kim. "Dynamic characteristics of suspension bridge hanger cables using digital image processing." NDT & E International 59 (2013): 25-33.

[8] Bazi, Yakoub, et al. "Differential evolution extreme learning machine for the classification of hyperspectral images." Geoscience and Remote Sensing Letters, IEEE 11.6 (2014): 1066-1070.

[9] Fan, Xiannian, Changhe Yuan, and Brandon M. Malone. "Tightening Bounds for Bayesian Network Structure Learning." AAAI. 2014.

[10] Fan, Xiannian, Brandon Malone, and Changhe Yuan. "Finding optimal Bayesian network structures with constraints learned from data." Proceedings of the 30th Annual Conference on Uncertainty in Artificial Intelligence (UAI-14). 2014.

[11] Despotović-Zrakić, Marijana, et al. "Scaffolding environment for adaptive e-learning through cloud computing." Journal of Educational Technology & Society 16.3 (2013): 301-314.

[12] Bullynck, Maarten. "Computing Primes (1929-1949): Transformations in the Early Days of Digital Computing." Annals of the History of Computing, IEEE 37.3 (2015): 44-54.

[13] Yang, Kan, and Xiaohua Jia. "An efficient and secure dynamic auditing protocol for data storage in cloud computing." Parallel and Distributed Systems, IEEE Transactions on 24.9 (2013): 1717-1726.

[14] Freitas, Leandro O., et al. "Applying pervasive computing in an architecture for homecare environments." Ubiquitous Intelligence & Computing and 9th International Conference on Autonomic & Trusted Computing (UIC/ATC), 2012 9th International Conference on. IEEE, 2012.

[15] Sathyanarayana, T. V., and L. Sheela. "Data security in cloud computing." Green Computing, Communication and Conservation of Energy (ICGCE), 2013 International Conference on. IEEE, 2013.

[16] Chugh, Sonam, and Sateesh Kumar Peddoju. "Access Control Based Data Security in Cloud Computing." International journal of Engineering Research and Applications 2.3 (2012): 2589-2593.

[17] Liu, Chang, et al. "An iterative hierarchical key exchange scheme for secure scheduling of big data applications in cloud computing." Trust, Security and Privacy in Computing and Communications (TrustCom), 2013 12th IEEE International Conference on. IEEE, 2013.

[18] Shahzad, Farrukh. "State-of-the-art survey on cloud computing security Challenges, approaches and solutions." Procedia Computer Science 37 (2014): 357-362.

[19] Chakravorty, Anjan, Tomasz Wlodarczyk, and Chunming Rong. "Privacy Preserving Data Analytics for Smart Homes." Security and Privacy Workshops (SPW), 2013 IEEE. IEEE, 2013.

[20] Patwal, Mayank, and Tanushri Mittal. "A Survey of Cryptographic based Security Algorithms for Cloud Computing." HCTL Open International Journal of Technology Innovations and Research 8 (2014): 2321-1814.